\documentclass[aps,prl,floatfix,twocolumn,showpacs]{revtex4}
\usepackage{epsfig}
\usepackage{amsmath}
\usepackage{amssymb}
\usepackage{amsfonts}
\usepackage{ dsfont }
\usepackage{ comment,color }

\newcommand{\mysection}[1]{{\it #1.}}

\newcommand{\beqa}{\begin{eqnarray}}
\newcommand{\eeqa}{\end{eqnarray}}

\begin{document}

\hsize\textwidth\columnwidth\hsize\csname@twocolumnfalse\endcsname

\title{
Proximity-induced Josephson $\pi$-Junctions in Topological Insulators}
\author{Constantin Schrade, A.~A. Zyuzin, Jelena Klinovaja, and Daniel Loss}

\affiliation{Department of Physics, University of Basel,
Klingelbergstrasse 82, CH-4056 Basel, Switzerland}

\date{\today}

\vskip1.5truecm
\begin{abstract}
We study two microscopic models of  topological insulators in contact with an $s$-wave superconductor. In the first model the superconductor and the topological insulator
are tunnel coupled via a layer of scalar and  of randomly oriented spin impurities. Here, we require that spin-flip tunneling dominates over  spin-conserving one. In the second model the tunnel coupling is realized by an array of single-level quantum dots with randomly oriented spins.
It is shown that the tunnel region forms a $\pi$-junction where the effective order parameter changes sign. Interestingly, due to the random spin orientation the effective descriptions of both models exhibit time-reversal symmetry. 
We then discuss how the proposed $\pi$-junctions support topological superconductivity  without  magnetic fields and can be used to generate and manipulate Kramers pairs of Majorana fermions by gates.
\end{abstract}

\pacs{74.50.+r, 71.10.Pm, 74.45.+c} 


\maketitle

\mysection{Introduction}
When two $s$-wave superconductors (SCs) are brought into contact via an insulator the 
energy of the system in equilibrium is minimized when the relative phase between the two
superconducting order parameters vanishes.
Interestingly, when the  insulator is doped with magnetic impurities, 
it was shown by theory \cite{bib:Sobyanin1977} and experiment \cite{bib:Moshchalkov} that 
spin-flip tunneling can induce an equilibrium ground state with a relative phase difference of $\pi$ between the superconducting order parameters, building up a so-called Josephson $\pi$-junction (J$\pi$J). 
It was predicted \cite{bib:Buzdin1990}  and  experimentally confirmed \cite{bib:Aarts2001} that a J$\pi$J can be generated by replacing the layer of magnetic impurities by a 
ferromagnetic metal. A J$\pi$J can also arise when two SCs are tunnel-coupled through 
an intermediate resonant state in the presence of strong Coulomb interactions \cite{bib:Kivelson1991}, as observed  in a system of two SCs  coupled by a  quantum dot (QD) occupied by a single electron \cite{bib:Kouwenhoven2006}.
In recent experiments \cite{bib:Yacoby2014,bib:Kouwenhoven2014,bib:Molenkamp2015} it was demonstrated that superconductivity can also be proximity-induced in the  
helical edge states of a topological insulator (TI) material \cite{bib:Kane2010,bib:Zhang2011,bib:Hankiewicz2013,bib:Pankratov1985, bib:Volkov1987, bib:Zhang2006,bib:Zhang2007, bib:Zhang2009, bib:Moler2012}
via coupling to an external $s$-wave SC. 
These experimental advances have also stimulated the theoretical interest in Josephson junctions based on TIs \cite{bib:Trauzettel2014,bib:Dolcini2014,bib:Hankiewicz2015}. 
Motivated by the existence of ordinary J$\pi$Js an important and immediate question is: 
Are there microscopic mechanisms allowing one to induce a superconducting order parameter in the helical edge states of the TI that is of {\it opposite} relative sign compared to the one of the external  $s$-wave SC, ideally without breaking time-reversal invariance (TRI)? In this work we answer this question in the affirmative. 

We propose two  setups involving TIs in which such a $\pi$-junction is shown to emerge. In the first setup the tunnel coupling is realized by a thin insulating layer of scalar and magnetic impurities with randomly oriented spins. Here we require that spin-flip tunneling dominates over normal tunneling. In the second setup the tunnel coupling is realized by an array of single-level QDs, each of which is occupied by a single spin with random orientation.
Critically, the random orientation of spins preserves TRI in an effective description. We note that both setups can be realized by combining the already existing experiments on proximity-inducing superconductivity solely in the edge states of a TI \cite{bib:Yacoby2014,bib:Kouwenhoven2014, bib:Molenkamp2015} and the experiments on J$\pi$Js in
SC-magnetic insulator-SC 
\cite{bib:Moshchalkov}
and SC-QD-SC devices \cite{bib:Kouwenhoven2006}.  We note that the same setup could be assembled in the framework of strip of stripes models \cite{Lebed,Kane_PRL,Stripes_PRL,Kane_PRB,Stripes_arxiv,yaroslav,QAHE} based on an array of coupled one-dimensional channels with spin-orbit interaction \cite{yaroslav}. As a striking consequence we find that the proposed models for proximity-induced J$\pi$Js in a TI provide an alternative approach to engineer  Kramers pairs of Majorana fermions (MFs) \cite{bib:Mele2013,bib:Berg2013,bib:Flensberg2014,bib:Oreg2014,bib:Law2012,bib:Nagaosa2013,bib:Law2014,bib:Tewari2014,bib:Loss12014,bib:Loss22014,bib:Trauzettel2011} easily movable by gates. Remarkably, no magnetic fields are needed.
More precisely we consider two TI samples that form a proximity-induced  J$\pi$Js with respect to one another and allow for tunneling between them in the finite space region, at the ends of which the MFs emerge.

\begin{figure}[t]  \centering
\includegraphics[width=1\linewidth] {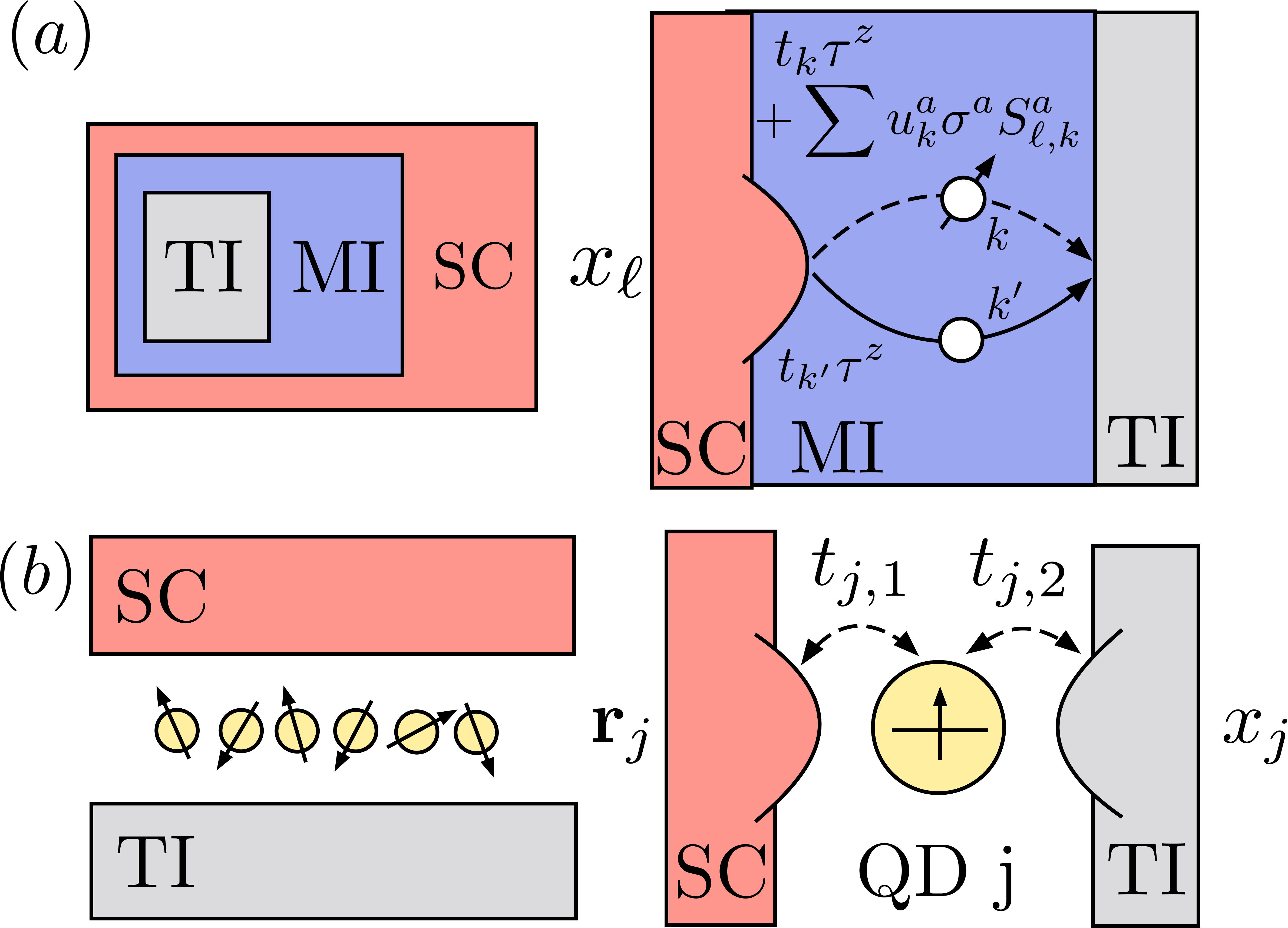}
\caption{(Color online) 
Setups to generate a proximity-induced Josephson $\pi$-junction in topological insulators (TIs). a) An $s$-wave SC (red) couples to a TI (grey) 
via an insulator doped with magnetic and scalar impurities (MI, magnetic insulator, blue). 
If the spin-flip tunneling rates are larger than the normal tunneling rates 
superconducting gaps with opposite sign are induced 
in the TI samples.  
b) Instead of the MI the SC is coupled to the TI via an array of single-level QDs in the
Coulomb blockade regime. 
The array of QDs is occupied with randomly oriented electron spins. 
}\label{fig:1}
\end{figure}

{\it Josephson junction models.}
In the first model we consider a bulk $s$-wave SC connected by a tunnel contact to the edge of a 2D TI, see Fig. \ref{fig:1}(a).
The Hamiltonian of the system is given by 

\begin{align}\label{H}
\text{H}_1 &= \text{H}_{\text{BCS}} + \text{H}_{\text{TI}}  \\ 
&
+\frac{1}{2} \int \mathrm{d}\textbf{r} \ \mathrm{d}x \  \left[\Psi^{\dag}(\mathbf{r})\cdot\mathrm{\bar T}_{1}
(\mathbf{r},x) \Phi(x) +\text{H.c.}\right],
\nonumber
\end{align}
with the tunneling matrix $\mathrm{\bar T}_{1}
(\mathbf{r},x)=\mathrm{T}_{1}
(\mathbf{r},x)(1+\tau^{z})/2-\mathrm{T}^{*}_{1}
(\mathbf{r},x)(1-\tau^{z})/2$.
Here, $\text{H}_{\text{BCS}}= (1/2)\int \mathrm{d}\textbf{r} \  \Psi^{\dag}(\mathbf{r})\cdot
[-(
\hbar^{2}\partial^{2}_{\textbf{r}}/2m + \mu
)\tau^{z}
-\Delta_{sc}\sigma^{y}\tau^{y}]\Psi(\mathbf{r}) +\text{H.c.}$  is the BCS Hamiltonian of the SC, $\mu$ being the chemical potential in the SC and $m$ being the electron mass, and 
$ \text{H}_{\mathrm{TI}}=(1/2)\int  \mathrm{d}x \  [\Phi^{\dagger}(x)\cdot(-i\hbar\upsilon_{F}\sigma^{z}\partial_{x}) \Phi(x) +\text{H.c.}]$ is the Hamiltonian of the TI edge with the Fermi velocity $\upsilon_{F}$. Without loss of generality, we assume that the superconducting order parameter $\Delta_{sc}$ is positive. The electron Nambu operator in the SC (TI) is given by $
\Psi(\textbf{r})
=(
\Psi_{\uparrow}(\textbf{r}), 
\Psi_{\downarrow}(\textbf{r}), 
\Psi^{\dagger}_{\uparrow}(\textbf{r}),
\Psi^{\dagger}_{\downarrow}(\textbf{r})
)$ [$
\Phi(x)=(
\text{R}(x), 
\text{L}(x), 
\text{R}^{\dagger}(x),
\text{L}^{\dagger}(x)
)$].
The Pauli matrices $\tau^{a}$ ($\sigma^{a}$) with $a= x,y,z$ act in particle-hole (spin) space
The slowly-varying spin-up right (spin-down left) mover fields $\text{R}(x)$ $\left[\text{L}(x)\right]$ are defined around the Fermi points $\pm k_{F}$ which in turn are determined by the position of the chemical potential $\mu_{\text{TI}}$ in the TI defined with respect to the Dirac point. 
The last term in Eq.~\eqref{H} describes the tunneling between points $\mathbf{r}$ of the SC and points $x$ on the edge of the TI. 
The interface between the SC and the TI is assumed to be rough, which means that the thinnest regions of the interface give the highest probability for electrons to tunnel between the SC and the TI. We model these thinnest regions located at points $x_{\ell}$ by point contacts. 
The tunnel contact 
between the SC and the TI is doped with scalar and magnetic impurities with randomly oriented spins $\textbf{S}_{\ell,k}=(S^x_{\ell,k},S^y_{\ell,k},S^z_{\ell,k})$. 
Here $\textbf{S}_{\ell, k}$ is the operator of the $k$-th localized spin close to the 
point contact $x_{\ell}$ on the TI sample.
The tunneling occurs via the virtual states of the scalar and magnetic impurities.
 The tunneling matrix amplitude is given by
\begin{align}
\label{tunneling1}
\text{T}_{1}(\textbf{r},x) &=\delta(\mathbf{r}-x\ \mathbf{e_{x}}) 
\\ &\ \ \times\sum_{\ell, k} \delta(x-x_{\ell})\left[ t_{k} 
+ \sum_{a=x,y,z} u^{a}_{k}\sigma^{a} S^{a}_{\ell, k}
\right].\nonumber
\end{align}
Here, $\mathbf{e_{x}}$ is a unit vector pointing along the TI edge written in terms of the coordinates of the SC. The normal (spin-flip) tunneling 
is parametrized by a complex amplitude $t_{k}$ ($u^{a}_{k}$) with
scalar impurities contributing to the amplitude $t_k$ only.
This model implies that there can be more than one 
magnetic or scalar impurity at the vicinity of the point contact.

In the second model we consider the coupling of a bulk $s$-wave SC to a 2D TI via an array of single-level
QDs, see Fig. 1(b). The Hamiltonian of the system is given by 
\begin{align}\label{H-system2}
&\text{H}_2 = \text{H}_{\text{BCS}} + \text{H}_{\text{TI}} + \text{H}_{\text{D}}  \\ 
&
+
\frac{1}{2}\sum_{j} 
\left[
 t_{j,1} \ D^{\dagger}_{j}\cdot \tau^{z}\Psi(\bold{r}_{j})
+
t_{j,2} \ D^{\dagger}_{j}\cdot \tau^{z}\Phi(x_{j})
+
\text{H.c.}
\right].
\nonumber
\end{align}
Here,
$
\text{H}_{\text{D}}= (1/2)
\sum_{j}
(-\epsilon_{j} \ D_{j}^{\dagger}\cdot\tau^{z}D_{j} +U_{j} \  n_{j,\uparrow}n_{j,\downarrow})
+ \text{H.c.}
$
is the Hamiltonian of an array of single-level QDs at energies $\epsilon_{j}>0$ and with amplitudes $U_{j}$
of the Coulomb interaction on the QDs and
 $
D_{j}=(
D_{j,\uparrow}, 
D_{j,\downarrow}, 
D^{\dagger}_{j,\uparrow},
D^{\dagger}_{j,\downarrow}
)$ is the electron Nambu operator on the $j$th QD. 
The occupation number operators for
spin-up and spin-down electrons on the $j$th QD  are given by
$n_{j,\uparrow}=D^{\dag}_{j,\uparrow}D_{j,\uparrow}$ 
and $n_{j,\downarrow}=D^{\dag}_{j,\downarrow}D_{j,\downarrow}$. 
Tunneling occurs at points $\bold{r}_{j}$ and $x_{j}$ 
on the SC and the TI, respectively,
and is described by
tunneling amplitudes $t_{j,1}$ and $t_{j,2}$. 

{\it Proximity-induced J$\pi$Js.}
We first discuss the model shown in Fig.~\ref{fig:1}(a) and described by Eqs.~(\ref{H}) and (\ref{tunneling1}). 
We neglect the inverse proximity effect due to magnetic impurities.  
By integrating out the degrees of freedom of the SC and including contributions up 
to second order in the tunneling amplitudes we see that the equation of motion 
for the Green's function $g(x,x')$ of the TI  in frequency space is given by 
\begin{equation}\label{eq4}
g^{-1}(x)\cdot g(x,x') = 
\delta(x-x') + \int \mathrm{d}x_{1}
\Sigma(x,x_{1})
\cdot g(x_{1},x')
\end{equation}
with $g^{-1}(x)= 
i\omega
+ i\hbar\upsilon_{F}\sigma^{z}\partial_{x}$ and 
$\omega$ the fermionic Matsubara frequency. 
In leading order,
the electron self-energy in the TI  is given by
\begin{eqnarray}\label{selfenergy}
\Sigma(x,x_1) =  
\int \mathrm{d}^{3}r \ \mathrm{d}^{3}r' \
\text{T}_{1}^{\dagger}(\textbf{r},x) \cdot G(\textbf{r}-\textbf{r}') \cdot
\text{T}_{1}(\textbf{r}',x_{1}).
\end{eqnarray} 
Here, $G(\textbf{r}-\textbf{r}')$ denotes the Green's function of the bare clean homogeneous three-dimensional SC defined by
$
G^{-1}(\textbf{r})\cdot G(\textbf{r}-\textbf{r}') = \delta(\textbf{r}-\textbf{r}')
$
with
$
G^{-1}(\mathbf{r})
= 
i\omega+
(
\hbar^{2}\partial^{2}_{\textbf{r}}/2m + \mu
)\tau^{z}
-
\Delta_{sc}\sigma^{y}\tau^{y}.
$
At vanishing relative distance a solution to this equation is given
\begin{eqnarray}
G(\boldsymbol{r}={0}) 
= 
\frac{-\pi
\nu
}
{
\sqrt{\omega^{2}
+\Delta_{sc}^{2} }
}
\left[\Delta_{sc}\sigma^{y}\tau^{y}+i\omega \right],~
\end{eqnarray}
with
$\nu=\frac{m  p_{F} }{2\pi^{2}}$ the normal-state density of states per spin and $p_{F}$ the Fermi momentum in the SC.
We adopt several assumptions to simplify Eq.~\eqref{selfenergy}. 
First, the distribution of impurities is 
assumed to be almost continuous and hence sums over impurities at discrete positions are replaced by integrals over impurity densities. Second, terms that are linear in the Pauli matrices $\sigma^{a}$ vanish
after averaging over the random orientation of the spins $\textbf{S}_{\ell, k}$.
Third, at some fixed $x_{\ell}$ tunneling contributions from points $x_{\ell'}$
for $\ell'\neq\ell$ can be neglected. The contribution of these terms to the effective Hamiltonian 
can be incorporated in the chemical potential \cite{bib:Kopnin}.
These assumptions imply that 
\begin{eqnarray}\label{minigap}\nonumber
\int  \mathrm{d}x_{1} \Sigma(x,x_{1})\cdot g(x_{1},x') \approx - \bigg[i\omega (\Gamma+\Gamma_{S}) \\
- \Delta_{sc}(\Gamma-\Gamma_{S})\sigma^{y}\tau^{y} \bigg]\cdot\frac{g(x,x')}{\sqrt{\omega^{2}+\Delta_{sc}^{2}}},
\end{eqnarray}
with the scattering rates
\begin{eqnarray}
 \label{GapGap}
\Gamma&=&\pi \nu n_{0} |\sum_{k} t_{k}|^{2}, \\
\Gamma_{S}&=& \pi \nu n_{S} S(S+1) \sum_{k,a}  \left|u^{a}_{k}\right|^{2}/3 . 
\label{rates}
\end{eqnarray}
Here, $n_{0}$ is the concentration of point contacts that allow for spin-conserved tunneling, while
$n_{S}$ is the concentration of point contacts that allow for spin-flip tunneling.
We note that 
$\langle \textbf{S}_{\ell, k}\textbf{S}_{\ell', k'} \rangle = S(S+1) \delta_{\ell\ell'} \delta_{kk'}$,
with $\langle...\rangle$ meaning the average over random spin directions and $S$ being the magnitude of the impurity spin. In particular, the average vanishes for different impurity spins. This implies that in the expression for 
the scattering rate $\Gamma_{S}$ terms $\propto u^{a}_{k}u^{a}_{k'}$ with $k\neq k'$ vanish as well.
The effective order parameter in the TI for  $\omega \ll \Delta_{sc}, \Gamma, \Gamma_S$ 
is given by 
\begin{equation}
\label{DeltaProximity}
\Delta_{\Gamma, \Gamma_S} \approx \Gamma - \Gamma_S.
\end{equation}
Interestingly, if $\Gamma_{S}>\Gamma$ the effective order parameter can become negative. 
Such a situation naturally emerges if
the tunnel contact contains a large number of magnetic and scalar impurities. At a particular point  $x_{\ell}$ the electron tunneling amplitude via some magnetic impurity $k$ is $t_{k} + \sum_{a}u^{a}_{k} \sigma^{a} S^{a}_{\ell, k} $ and $t_{k'}$ for some scalar impurity $k'$. 
We assume that $|t_{k}|\approx |t_{k'}|$, while generally their relative sign is random. Thus,
for many impurities the normal tunneling contributions in Eq.~\eqref{GapGap} destructively interfere, so that $\sum_k t_k \approx 0$. As a result, $\Gamma_{S}>\Gamma$  can be realized and $\Delta_{\Gamma, \Gamma_S} $ becomes negative.

Next we discuss the model of an $s$-wave SC coupled to a 2D TI via an array of QDs, as depicted in Fig.~\ref{fig:1}(b)
and described by Eq.~(\ref{H-system2}).
We will work in the Coulomb blockade regime.
Thus, we assume singly occupied QDs with the electron spin on the QDs being randomly oriented. In the limit of small tunneling amplitudes that
couple the SC and the TI to the QD we use a Schrieffer-Wolff transformation \cite{bib:Wolff1966} to map the Hamiltonian $\text{H}_2$ as given in Eq.~\eqref{H-system2} onto a Hamiltonian  $\text{H}_1$ of the form as given in Eq.~\eqref{H} with $t_{k}\equiv 0$. The physical interpretation is that due to the large Coulomb interactions 
on the QDs only spin-flip tunneling of electrons through the dots is allowed \cite{bib:Kivelson1991}. 
From the discussion of the first model we can conclude again that 
$\Gamma_{S}>0$, while
$\Gamma\approx 0$.

Thus, we see that in both models we obtain a J$\pi$J in the tunneling region, {\it i.e.}, the proximity-induced effective superconducting order parameter in the helical edge states of the TI  assumes the opposite  sign compared to the one of the  SC. 

{\it Kramers pairs of Majorana fermions.} 
In a TI proximity-coupled to an $s$-wave SC magnetic perturbations
can be used to induce MFs \cite{bib:Fu2008}. However, in realistic scenarios the use of  magnetic fields
should be avoided since it acts detrimental on 
 superconductivity and
it is indeed not a necessary ingredient: 
In the absence of it Kramers pairs of MFs can emerge in nanowire systems that are coupled to unconventional SCs \cite{bib:Law2012,bib:Nagaosa2013,bib:Law2014,bib:Tewari2014,bib:Mele2013}. Also setups using conventional SCs in proximity to nanowires \cite{bib:Berg2013, bib:Flensberg2014, bib:Oreg2014, bib:Loss12014} and in 2D \cite{bib:Loss22014} and to 3D TIs \cite{bib:Trauzettel2011} were proposed. 
In particular, it was predicted that Kramers pairs of MFs appear due to J$\pi$Js
in nanowires \cite{bib:Berg2013,bib:Oreg2014,bib:Mele2013} or in 3D TI films \cite{bib:Trauzettel2011}. 
In this section we make use of the J$\pi$J models introduced above  and propose two  setups (labeled by $N=1,2$)
that
host  Kramers pairs of MFs based on two 2D TIs. 
As a major advantage both setups are 
accessible by current experimental techniques
in TIs \cite{bib:Yacoby2014,bib:Kouwenhoven2014, bib:Molenkamp2015} and 
in J$\pi$Js based on magnetic insulators \cite{bib:Moshchalkov} and QDs \cite{bib:Kouwenhoven2006}. 
Also, to reveal the non-abelian statistics the pairs
can easily be moved by tuning a
tunnel barrier between the TIs.

\begin{figure}[t] \centering
\includegraphics[width=1\linewidth] {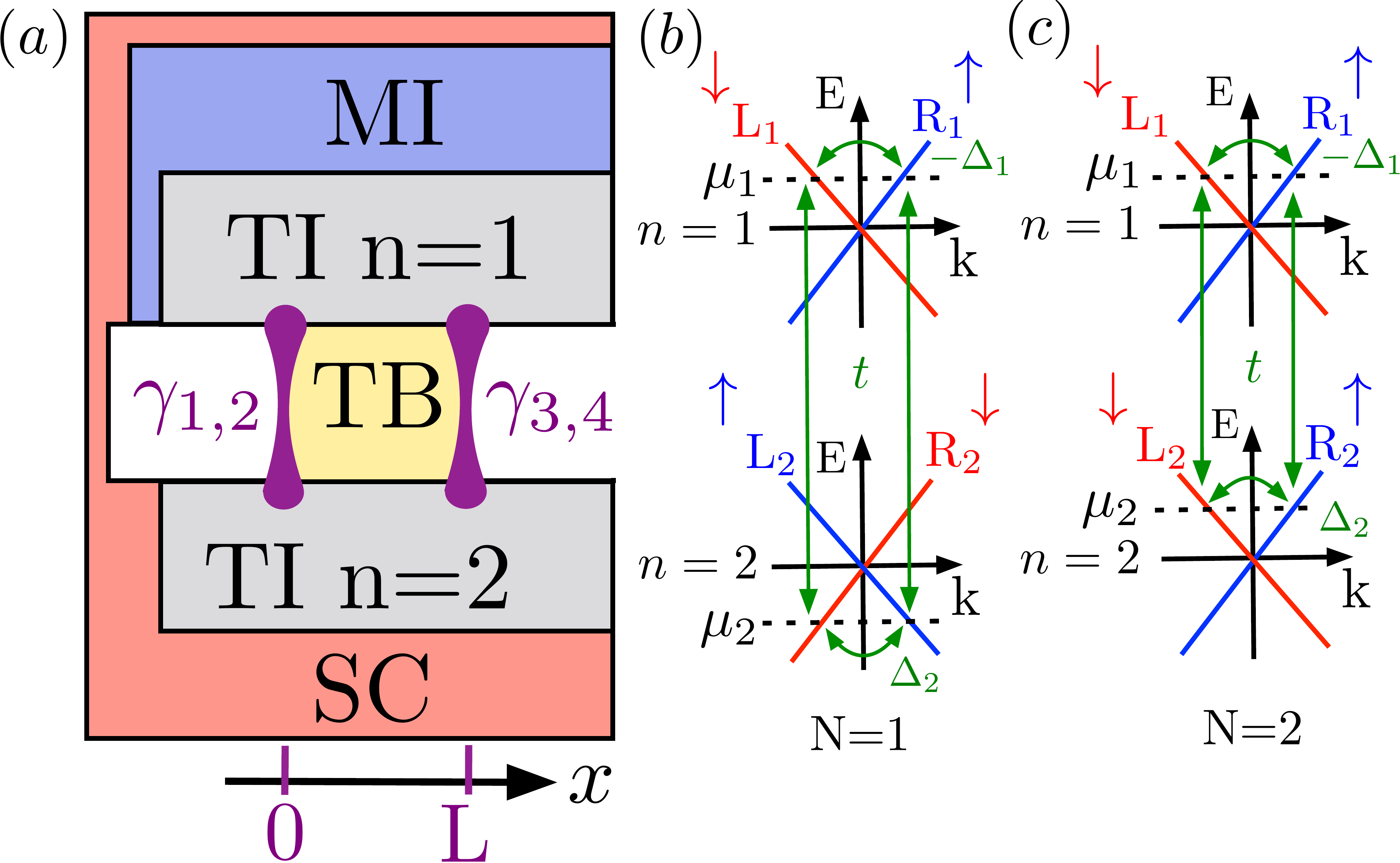}
\caption{(Color online)
(a) Setup hosting Kramers pairs of MFs. Two TIs (grey rectangles) are placed on top of an underlying $s$-wave SC (red) such that proximity superconductivity is induced in both TIs. Importantly, the first TI is coupled through a magnetic insulator (MI, blue) resulting in the J$\pi$J.
The tunnel barrier (TB, yellow) between the edges of two TIs extends from $x=0$ to $x=L$. One Kramers pair of MFs $\gamma_{1,2}$ [$\gamma_{3,4}$] (purple) is
localized at the $x=0$ [$x=L$] end of the TB and can be manipulated by tuning the length L of the TB. The spectrum of two pairs of TI edge modes is considered in (b) for the first setup and in (c) for the second  setup.
(b) Edge modes of the same TI are coupled by proximity-induced pairing amplitudes $\Delta_{1}<0$ and $\Delta_{2}>0$, {\it resp.} The chemical potentials are opposite for the two TIs, $\mu_{1}=-\mu_{2}$. 
The helicities of the edge states are opposite (indicated by the coloring in red and blue). 
The tunneling ($t$) couples a right-moving state in the first TI  to a left-moving state in the second TI, and vice versa. (c)  The two TIs have the same chemical potential, $\mu_{1}=\mu_{2}$, and the same helicities. The tunneling ($t$) couples a right-[left-] moving state in the first TI to a right-[left-] moving state in the second TI. 
}\label{fig2}
\end{figure}

We consider two TIs labeled by $n=1,2$.
 In the first (second) setup, edge states are of opposite (same) helicity and the chemical potentials are tuned to be opposite (to be the same) with $\mu_{1}=-\mu_{2}$ ($\mu_{1}=\mu_{2}$),
as illustrated in Fig.~\ref{fig2}b (Fig.~\ref{fig2}c).  
Both TIs are brought into proximity to an $s$-wave SC. 
In the the first TI, the tunnel contact is doped with scalar and magnetic impurities with randomly oriented spins or, equivalently, an array of QDs with randomly oriented 
spins is used. 
As shown above, a $\pi$-junction emerges and the proximity gap in the first TI acquires the opposite sign to the bulk SC, $-\Delta_1<0$. The tunnel contact between the SC and the second TI  does not contain a spin-flip source. Thus, the corresponding order parameter is of the same sign 
as in the SC, $\Delta_2>0$. 
The induced superconductivity in the $n$th TI of the $N$th setup is described by the Hamiltonian
\begin{equation}
\begin{split}
 H_{sc,n}^{(N)}
=(-1)^{(N-1)(n-1)} 
\frac{\Delta_{n}}{2}
\int \mathrm{d}x 
\left[
\text{L}^{\dag}_{n}\text{R}^{\dag}_{n}-\text{R}^{\dag}_{n}\text{L}^{\dag}_{n}
+
\text{H.c.}
\right]
\end{split}\label{HSC}
\end{equation}
in the basis $ \Phi_{n}(x) =({\text{R}}_{n}(x), {\text{L}}_{n}(x),{\text{R}}_{n}^{\dagger}(x), {\text{L}}_{n}^{\dagger}(x))$.
For the first (second) TI  of the first setup we have introduced slowly-varying spin-up (spin-down) right-mover  $\text{R}_{1}(x)$ [$\text{R}_{2}(x)$] and spin-down (spin-up) left-mover $\text{L}_{1}(x)$ [$\text{L}_{2}(x)$] fields defined around the Fermi points $\pm k_{F}$. In the second setup $\text{R}_{2}(x)$ [$\text{L}_{2}(x)$] is the spin up (spin down) mode, see Fig. \ref{fig2}.

The two TIs are coupled via a tunnel barrier placed in the region $0\leqslant x \leqslant L$ as shown in Fig.~\ref{fig2}(a) Neglecting the fast-oscillating terms \cite{bib:Klinovaja2012}, we find that the tunneling Hamiltonian in the first setup is given by 
\begin{align}
&\text{H}_{t}^{(1)}= \frac{t}{2} \int_{0}^{L} \mathrm{d}x 
[
e^{i\phi}
\left(
\text{R}^{\dagger}_{2}\text{L}_{1}
-
\text{L}_{1}\text{R}^{\dagger}_{2}
\right)\\
&\hspace{80pt}+e^{-i\phi}
\left(
\text{L}_{2}^{\dagger}\text{R}_{1}
-
\text{R}_{1}
\text{L}_{2}^{\dagger}
\right)
+\text{H.c.}
] , \nonumber
\end{align}
while in the second setup by
\begin{align}
&\text{H}_{t}^{(2)}= \frac{t}{2} 
\int_{0}^{L} \mathrm{d}x 
[
e^{i\phi}
\left(
\text{R}^{\dagger}_{2}\text{R}_{1}
-
\text{R}_{1}\text{R}^{\dagger}_{2}
\right)\\
&\hspace{80pt}+e^{-i\phi}
\left(
\text{L}^{\dagger}_{2}\text{L}_{1}
-
\text{L}_{1}
\text{L}^{\dagger}_{2}
\right)
+\text{H.c.}
].\nonumber
\end{align}
Here, $t$ ($\phi$) is the tunneling amplitude (phase) between two TIs. 
The choice of $\phi$ ensures  TRI. 
The total Hamiltonian is $H^{(N)}=\sum_{n} (H_{TI,n}+ H_{sc,n}^{(N)}) +\text{H}_{t}^{(N)}$, where the kinetic part $H_{TI,n}\equiv H_{TI}$ is identical for both TIs  and was introduced in Eq. (\ref{H}).
The tunneling phase can be removed from the total Hamiltonian by a suitable gauge transformation \cite{bib:supplemental}.
In both setups,  we find that a topological phase transition accompanied by the bulk gap closing and reopening occurs at the point
\begin{equation}
t=\sqrt{\Delta_{1}\Delta_{2}}.
\end{equation}
In the second setup, there is an additional constraint $\Delta_{1}\neq\Delta_{2}$. 
If $t>\sqrt{\Delta_{1}\Delta_{2}}$, there is one Kramers pair of 
MFs  localized at the interfaces at $x=0$ and one at $x=L$ \cite{bib:Klinovaja2012}.
The localization lengths are inversely proportional to the gaps opened at the Fermi points \cite{bib:supplemental}. Thus, in regions with no tunnel coupling between the TIs the localization 
lengths are the superconducting coherence lengths 
$\xi_{n}=\hbar\upsilon_{F}/\Delta_{n}$,
while in regions with $t>\sqrt{\Delta_{1}\Delta_{2}}$ they are given by 
\begin{align}
\xi^{(1)}&=2\hbar\upsilon_{F}/(\sqrt{(\Delta_{1}-\Delta_{2})^{2}+4t^{2}}-\Delta_{1}-\Delta_{2}) \\
\xi^{(2)}_{\pm}&=
\frac{
2\hbar\upsilon_{F}}{
|\Delta_{1}-\Delta_{2}|\pm
\Re
\sqrt{(\Delta_{1}+\Delta_{2})^{2}-4t^{2}}
}.
\end{align}
Superscript $(1)$ [(2)] corresponds to first [second] setup and $\Re$ means the real part 
of a complex number. In both setups  we assume that the length of the tunnel barrier $L$ is much longer
than the localization lengths
$\xi^{(1)}_{\text{max}}\equiv\text{max}\{\xi_{1},\xi_{2},\xi^{(1)}\}$ 
($\xi^{(2)}_{\text{max}}=\text{max}\{\xi_{2},\xi^{(2)}_{-}\}$ for $\Delta_{1}>\Delta_{2}$ and $\xi^{(2)}_{\text{max}}=\text{max}\{\xi_{1},\xi^{(2)}_{-}\}$ for $\Delta_{1}<\Delta_{2}$).
Hence the wavefunctions of the MFs localized
at the two different interfaces do not overlap and can be considered as independent.  
If $L$ is comparable or shorter than the localization length of the MFs they hybridize into two complex fermionic state whose energies are non-zero in general \cite{bib:Rainis2013,bib:Zyuzin2013}.
Tuning $L$
by underlying gates allows one to arbitrarily control 
the position of the MFs along the TI edges. 
To give a numerical estimate of the localization length we assume 
that
the induced gaps are given by $\Delta_{1}=0.1$ meV, $\Delta_{2}=0.2$ meV, 
and the tunnel coupling is set to $t=0.2$ meV. In a 
InAs/GaSb (HgTe/CdTe) TI the Fermi velocity is given by $\upsilon_F=4.6\times 10^{4}\ \text{m}\, \text{s}^{-1}$ \cite{bib:Kouwenhoven2014}
($5.5\times 10^{5}\ \text{m}\, \text{s}^{-1}$ \cite{bib:Zhang2011}). This gives a localization length of the order of $0.5\ \mu$m ($5\ \mu$m).
Finally, we emphasize here that the opposite relative sign in front of the proximity induced gaps of the two edges in Eq.\eqref{HSC} 
is an important ingredient to generate the Kramers pair of MFs.
If this relative sign  is the same there exists no topological phase. For illustrative phase diagrams, see \cite{bib:supplemental}.

{\it Conclusions.}
We have proposed and studied two setups to realize a proximity-induced J$\pi$J in a TI
in the presence of TRI.
Both setups rely on the tunnel coupling of a TI sample to an $s$-wave bulk SC
either via a layer of scalar and magnetic impurities with randomly oriented spins 
or via an array of QDs each of which is occupied by a randomly oriented spin. 
We have seen that if in either case spin-flip tunneling dominates over normal tunneling 
a $\pi$-junction emerges.
The randomly oriented spins 
ensure that 
there is effectively no breaking of TRI. 
As an application we have demonstrated how such  proximity-induced $\pi$-junctions  can be used 
to generate and manipulate Kramers pairs of MFs in edge states of tunnel-coupled TIs. 

{\it Acknowledgments.}
We acknowledge support  from the Swiss NSF and NCCR QSIT.

\renewcommand{\figurename}{Figure S\!\!}
\setcounter{figure}{0}
\renewcommand\thesection{S\arabic{section}} 
\renewcommand\thesubsection{S\arabic{section}.\arabic{subsection}}
\renewcommand\bibnumfmt[1]{S#1.} 
\renewcommand\theequation{S\arabic{equation}}
\setcounter{equation}{0}

\renewcommand{\footnote}[1]{\footnotemark\footnotetext{#1}}
\renewcommand{\thefootnote}{\alph{footnote}}
\setcounter{footnote}{0}

\def\hksqrt{\mathpalette\DHLhksqrt}
\def\DHLhksqrt#1#2{\setbox0=\hbox{$#1\sqrt{#2\,}$}\dimen0=\ht0
\advance\dimen0-0.2\ht0
\setbox2=\hbox{\vrule height\ht0 depth -\dimen0}%
{\box0\lower0.4pt\box2}}

\newpage

\onecolumngrid

\begin{center}
\large{\bf Supplemental Material to `Proximity-induced Josephson $\pi$-junctions in topological insulators'}
\vspace{-0.3cm}
\end{center}
\begin{center}
Constantin Schrade, A.A. Zyuzin, Jelena Klinovaja, and Daniel Loss\\
{\it Department of Physics, University of Basel,
Klingelbergstrasse 82, CH-4056 Basel, Switzerland}
\end{center}

\twocolumngrid

In the Supplemental Material, we derive the MF wavefunctions for two models introduced in the main text.

\subsection{Energy spectrum}

We find that the bulk spectrum of the Hamiltonian $H^{(1)}$ from the main text is given by
\begin{equation} 
E_{1, s,\pm}(k) = s \left[ (\hbar\upsilon_{F}k)^{2} +\Big(\Delta_{+}\pm\sqrt{\Delta_{-}^{2}+t^{2}}\Big)^{2} \right]^{1/2},
\end{equation}
where $k$ is the momentum in the TI, and $s=\pm1$. 
Similarly the bulk spectrum of the Hamiltonian $H^{(2)}$ from the main text is given by 
\begin{equation}  
E_{2, s,\pm}(k) =s \left[(\hbar\upsilon_{F}k)^{2} +\Delta_{+}^{2}+\Delta_{-}^{2}+t^{2}\pm 2\sqrt{W(k)}\right]^{1/2}
\end{equation} 
with $W(k)= (\hbar\upsilon_{F}k)^{2}t^{2}+\Delta_{+}^{2}(\Delta_{-}^{2}+t^{2})$.
Here, we also introduced the notations $\Delta_{\pm} = (\Delta_{1}\pm\Delta_{2})/2$.
 Both spectra 
$E_{1, s,\pm}(k)$ and $E_{2, s,\pm}(k)$ are twofold degenerate as expected for time-reversal invariant systems.

We find that the spectrum is gapless at $k=0$ if
\begin{equation}  
t=\sqrt{\Delta_{1}\Delta_{2}} ,
\label{TT}
\end{equation}
and is gapped otherwise. Here, for setup $N=2$ we need the additional condition that $\Delta_{1}\neq\Delta_{2}$.
Also the spectral gap for the setup $N=2$ closes at some finite momentum if $t>\Delta_{1}=\Delta_{2}$.
We now assume that $\Delta_{1}\neq\Delta_{2}$ and 
confirm that Eq. (\ref{TT}) defines a topological phase transition. This means that there should be MFs localized at the boundary between two space regions with $t>\sqrt{\Delta_{1}\Delta_{2}} $
and $t<\sqrt{\Delta_{1}\Delta_{2}} $.

\subsection{Wavefunctions and localization lengths of MFs}
The operator defining a MF, which is a zero-energy bound state,  is generally given by
$\gamma_{j}^{(N)}\equiv(\gamma_{j}^{(N)})^\dagger= \sum_{n=1,2}\int \mathrm{d}x \ \psi_{n,j}^{(N)}(x)\cdot{\Phi}_{n}(x)$ with the wavefunction (vector)
\begin{align}
(\psi_{n,j}^{(N)})^{T}(x)=
\begin{pmatrix}
f_{n,j}^{(N)}(x) \\
g_{n,j}^{(N)}(x)\\
(f_{n,j}^{(N)})^{*}(x)\\
(g_{n,j}^{(N)})^{*}(x)\\
\end{pmatrix}
\end{align}
for some complex-valued functions $f_{n,j}^{(N)}(x)$ and $g_{n,j}^{(N)}(x)$. The index $j=1,2$ distinguishes between two MFs belonging to the same Kramers pair.
The form of these functions is different for different setups.

Without loss of generality, we focus below on the left interfaces at which the tunneling amplitude jumps from $t=0$ at $x<0$ to $t=t_{0}>\sqrt{\Delta_{1}\Delta_{2}}$ for $x>0$.

{\it First setup.} We find that for the first setup 
the interface hosts a Kramers pair of MFs given by

\begin{figure*}[t] \centering
\includegraphics[width=0.6\linewidth] {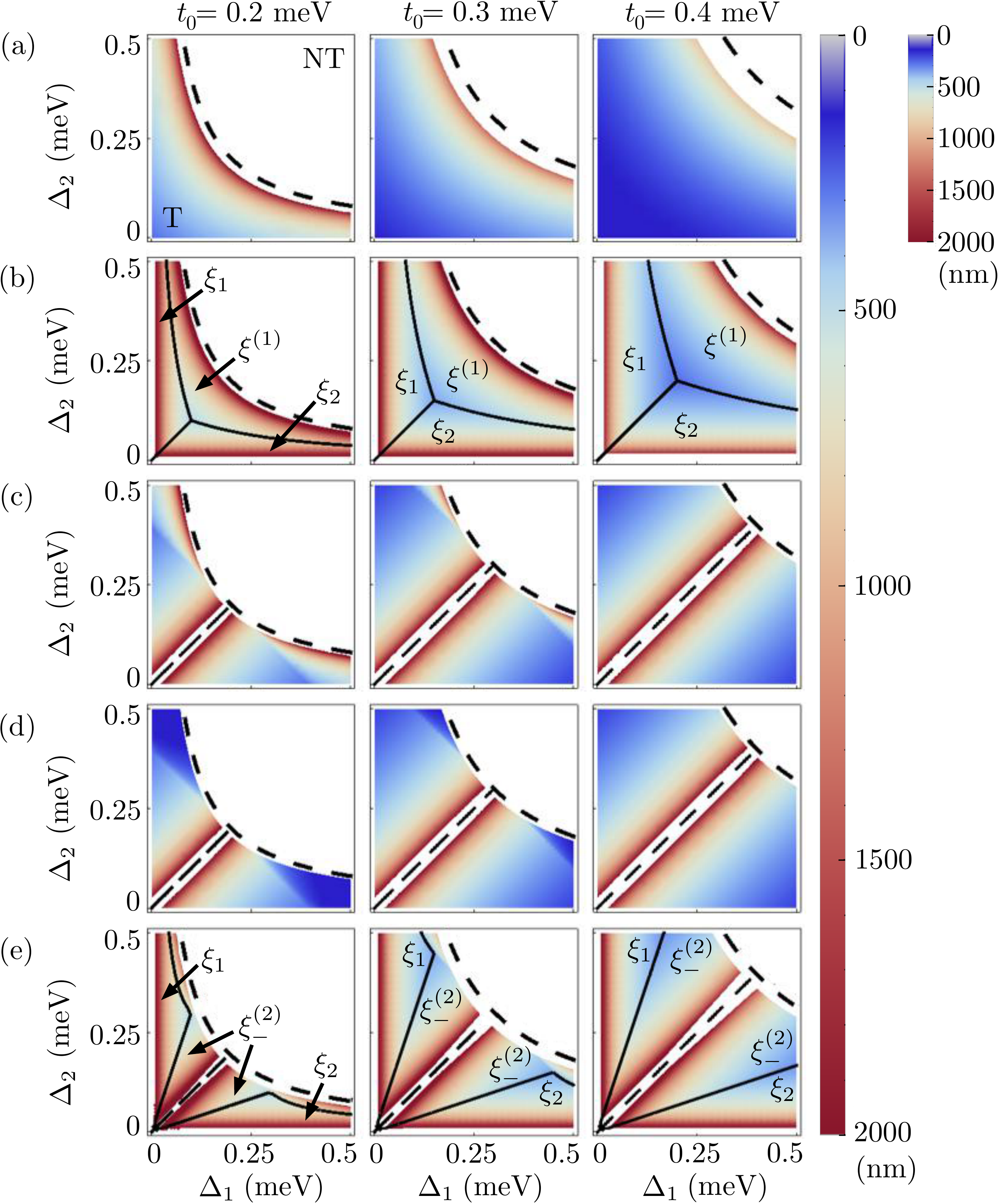}
\caption{(Color online)
(a) Phase diagrams and color scale plots of the localization length $\xi^{(1)}$ for $x>0$ of the Kramers pair of MFs in the first setup as 
a function of the superconducting gap parameters $\Delta_{1,2}$ and the tunneling amplitude $t_{0}$. 
Here, $\xi^{(1)}$ increases from blue, through yellow, to red; $v_F=4.6\times 10^{4}\ \text{m}\, \text{s}^{-1}$ in an InAs/GaSb TI \cite{bib:Kouwenhoven2014}.
The curve $t_{0}=\sqrt{\Delta_{1}\Delta_{2}}$ (dashed) seperates the topological phase (T, colored) and the non-topological phase (NT, uncolored). At the phase boundary the localization length $\xi^{(1)}$ is divergent.
For $x<0$ the localization lengths are given by the superconducting coherence lengths $\xi_{1,2}=\hbar\upsilon_{F}/\Delta_{1,2}$. (b) Same as in (a) but for 
$\xi^{(1)}_{\text{max}}\equiv\text{max}\{\xi_{1},\xi_{2},\xi^{(1)}\}$. The curves $\xi_{1,2}=\xi^{(1)}$
and $\xi_{1}=\xi_{2}$ (solid) seperate regions where $\xi^{(1)}_{\text{max}}$ is given respectively by 
$\xi_{1},\xi_{2}$ or $\xi^{(1)}$. 
(c) Same as in (a) but for the localization length $\xi^{(2)}_{-}$ for $x>0$ 
in the second setup. Along the line $\Delta_{1}=\Delta_{2}$ (dashed) the localization lengths $\xi^{(2)}_{\pm}$ are divergent.
(d) Same as in (c) but for the localization length $\xi^{(2)}_{+}$.
(e) Same as in (d) but for $\xi^{(2)}_{\text{max}}=\text{max}\{\xi_{2},\xi^{(2)}_{-}\}$ if $\Delta_{1}>\Delta_{2}$ and $\xi^{(2)}_{\text{max}}=\text{max}\{\xi_{1},\xi^{(2)}_{-}\}$ if $\Delta_{1}<\Delta_{2}$. The curves $\xi_{1,2}=\xi^{(2)}_{-}$ (solid) seperate regions where $\xi^{(2)}_{\text{max}}$ is given respectively by 
$\xi_{1},\xi_{2}$ or $\xi^{(2)}_{-}$. 
}\label{fig_sup}
\end{figure*}

\begin{widetext}
\begin{equation}
\begin{split}
-i
f_{n,1}^{(1)} &=
(g_{n,1}^{(1)})^{*}
=
\begin{cases}
\Big(\delta_{n1}
\frac{\sqrt{\Delta^{2}_{-}+t_{0}^{2}}-\Delta_{-}}{t_{0}}
 e^{ik_{F}x}
+\delta_{n2}e^{-ik_{F}x}\Big)e^{\frac{i\phi}{2}}e^{-x/\xi^{(1)}} & \mbox{if } x>0 
\\
\Big(
\delta_{n1}
\frac{\sqrt{\Delta^{2}_{-}+t_{0}^{2}}-\Delta_{-}}{t_{0}}
 e^{ik_{F}x} e^{x/\xi_{1}} + \delta_{n2} e^{-ik_{F}x}e^{x/\xi_{2}}\Big)e^{\frac{i\phi}{2}}  & \mbox{if } x<0 
\end{cases},\\
f_{n,2}^{(1)}&=
(-1)^{n}(g_{n,1}^{(1)})^{*},
\quad
g_{n,2}^{(1)} =(-1)^{n-1}
(f_{n,1}^{(1)})^{*}
\end{split}
\end{equation}
\end{widetext}
with the localization lengths given by
\begin{equation}
\begin{split}
&\xi^{(1)}= 
\hbar\upsilon_{F}/
(
\sqrt{\Delta_{-}^{2}+t_{0}^{2}}-\Delta_{+}
)
\\
&\xi_{n}=\hbar\upsilon_{F}/\Delta_{n}.
\end{split}
\end{equation}
In Fig. S\ref{fig_sup}(a) the localization length $\xi^{(1)}$ is plotted for different valus of $t_{0}$ in color scale versus $\Delta_{1}$ 
and $\Delta_{2}$.
Note that the solutions for given $N$ are orthogonal, $\psi_{n,1}^{(N)}\cdot (\psi_{n,2}^{(N)})^{T}=0$, thus forming a Kramers pair.
The localization length of the MF is given by $\xi_{\text{max}}^{(1)}=\text{max}\{\xi^{(1)},\xi_{1},\xi_{2}\}$ 
and is plotted in Fig. S\ref{fig_sup}(b).

{\it Second setup.} 
The interface at $x=0$ of the second setup also hosts a Kramers pair of MFs.
For $\Delta_{-}>0$ the Kramers pair of MFs is given by
\begin{widetext}
\begin{equation}
\begin{split}
f_{n,1}^{(2)} &=
 e^{ik_{F}x}
\begin{cases}
i\delta_{n2}\left(
\frac{
\Delta_{+}-\sqrt{\Delta^{2}_{+}-t_{0}^{2}}
}{2\sqrt{\Delta^{2}_{+}-t_{0}^{2}}}
e^{-x/\xi^{(2)}_{+}}-
\frac{
\Delta_{+}+\sqrt{\Delta^{2}_{+}-t_{0}^{2}}
}{2\sqrt{\Delta^{2}_{+}-t_{0}^{2}}}
e^{-x/\xi^{(2)}_{-}}\right)e^{i\frac{\phi}{2}}
-
\frac{
\delta_{n1}
(e^{-x/\xi^{(2)}_{-}}-e^{-x/\xi^{(2)}_{+}})t_{0}e^{-i\frac{\phi}{2}}
}{
2\sqrt{\Delta^{2}_{+}-t_{0}^{2}}
}
 & \mbox{if } x\geq0, \  \Delta_{+}>t_{0}
\\
-
\left(
\delta_{n1}  \frac{t_{0} x}{\hbar\upsilon_{F}}  e^{-i\frac{\phi}{2}} 
+i\delta_{n2} 
(1+\frac{t_{0}x}{\hbar\upsilon_{F}})e^{i\frac{\phi}{2}}
\right)e^{-x/\xi^{(2)}_{\pm}}
 & \mbox{if } x\geq0, \  \Delta_{+}=t_{0}
\\
-
\left(
\delta_{n1}
\frac{t_{0}
\sin(k^{(2)}x)
}{
\sqrt{t_{0}^{2}-\Delta^{2}_{+}}
}
e^{-i\frac{\phi}{2}}
+i\delta_{n2}\left(
\frac{
\Delta_{+}\sin(k^{(2)}x)
}{\sqrt{t_{0}^{2}-\Delta^{2}_{+}}}
+
\cos(k^{(2)}x)
\right)
e^{i\frac{\phi}{2}}
\right)e^{-x/\xi^{(2)}_{\pm}}
 & \mbox{if } x\geq0, \  \Delta_{+}<t_{0}
\\
-i
\delta_{n2} \ e^{i\frac{\phi}{2}}
e^{x/\xi_{2}}
 & \mbox{if } x<0 
\end{cases},\\
g_{n,1}^{(2)}&=i(f_{n,1}^{(2)})^{*}, \quad 
f_{n,2} ^{(2)}=-(g_{n,1}^{(2)})^{*}, \quad
g_{n,2}^{(2)}=(f_{n,1}^{(2)})^{*}\, ,
\end{split}
\end{equation}
\end{widetext}
while for $\Delta_{-}<0$ it is given by 
\begin{widetext}
\begin{equation}
\begin{split}
f_{n,1}^{(2)} &=
e^{ik_{F}x}
\begin{cases}
\frac{
i
\delta_{n2}
\left(
e^{-x/\xi^{(2)}_{+}}-e^{-x/\xi^{(2)}_{-}}
\right)t_{0}e^{i\frac{\phi}{2}}
}{
2\sqrt{\Delta^{2}_{+}-t_{0}^{2}}
}-\delta_{n1}\left(
\frac{
\Delta_{+}-\sqrt{\Delta^{2}_{+}-t_{0}^{2}}
}{2\sqrt{\Delta^{2}_{+}-t_{0}^{2}}}
e^{-x/\xi^{(2)}_{+}}-
\frac{
\Delta_{+}+\sqrt{\Delta^{2}_{+}-t_{0}^{2}}
}{2\sqrt{\Delta^{2}_{+}-t_{0}^{2}}}
e^{-x/\xi^{(2)}_{-}}\right)e^{-i\frac{\phi}{2}}
 & \mbox{if } x\geq0, \ \Delta_{+}>t_{0}
\\
\left(
\delta_{n1}
(1+\frac{t_{0} x}{\hbar\upsilon_{F}})
e^{-i\frac{\phi}{2}}
-i
\delta_{n2} \ 
\frac{t_{0} x}{\hbar\upsilon_{F}} \ e^{i\frac{\phi}{2}}
\right)e^{-x/\xi^{(2)}_{\pm}}
 & \mbox{if } x\geq0, \ \Delta_{+}=t_{0}
\\
\left(
\delta_{n1}
\left(
\frac{
\Delta_{+}\sin(k^{(2)}x)
}
{\sqrt{t_{0}^{2}-\Delta^{2}_{+}}
}
+
\cos(k^{(2)}x)
\right)
e^{-i\frac{\phi}{2}}
-i
\delta_{n2}
\frac{
t_{0}\sin(k^{(2)}x)
}
{
\sqrt{t_{0}^{2}-\Delta^{2}_{+}}
}
e^{i\frac{\phi}{2}}
\right)
e^{-x/\xi^{(2)}_{\pm}}
 & \mbox{if } x\geq0, \ \Delta_{+}<t_{0}
\\
\delta_{n1} \ e^{-i\frac{\phi}{2}}
e^{x/\xi_{1}}
 & \mbox{if } x<0 
\end{cases}\,\,,\\
g_{n,1}^{(2)}&=i(f_{n,1}^{(2)})^{*},\quad
f_{n,2} ^{(2)}=-(g_{n,1}^{(2)})^{*},\quad
g_{n,2}^{(2)}=(f_{n,1}^{(2)})^{*} .
\end{split}
\end{equation}
\end{widetext}
Again, the two wavefunctions for given $N$ are orthogonal and thus forming a Kramers pair.
We have defined the wavenumber
\begin{equation}
k^{(2)}=\sqrt{|\Delta^{2}_{+}-t_{0}^{2}|}/\hbar\upsilon_{F}
\end{equation}
and the localization lengths
\begin{equation}
\begin{split}
\xi^{(2)}_{\pm}
=
\frac{
\hbar\upsilon_{F}}{
|\Delta_{-}|\pm
\Re\sqrt{\Delta^{2}_{+}-t_{0}^{2}}}\, .
\end{split}
\end{equation}
For different values of $t_{0}$ the localization lengths $\xi^{(2)}_{\pm}$ 
are displayed in Fig. S\ref{fig_sup}(c) and (d) in a color scale plot as a function of $\Delta_{1}$ and $\Delta_{2}$.
For $t_{0}\leq\Delta_{+}$ the spectral gap closes at zero momentum while for $t_{0}>\Delta_{+}$
it closes at some finite momentum. 
For $\Delta_{-}>0$ ($\Delta_{-}<0$) the localization length of the MF is given by $\xi_{\text{max}}^{(2)}=\text{max}\{\xi_{2},\xi^{(2)}_{-}\}$ ($\xi_{\text{max}}^{(2)}=\text{max}\{\xi_{1},\xi^{(2)}_{-}\}$) and is plotted in Fig. S\ref{fig_sup}(e). 
The fast-oscillating factors have explicitly been restored in the wavefunctions. 
We see that for $\Delta_{1}=\Delta_{2}$ the MF wavefunction is delocalized.
In the limit when $L\gg\xi_{\text{max}}^{(N)}$ the interfaces at $x=0$ and $x=L$ can be considered as 
independent and a calculation of the MF wavefunctions at $x=L$ can be performed analogously.

Finally for setup $N$ the tunneling phase $\phi$ 
can be absorbed into a redefinition of the electron operators by the gauge transformations
\begin{equation}
\begin{split}
\text{R}_{n}&\mapsto \exp\left((-1)^{n(N-1)}\frac{i\phi}{2}\right)\text{R}_{n}\\
\text{L}_{n}&\mapsto \exp\left(-(-1)^{n(N-1)}\frac{i\phi}{2}\right)\text{L}_{n}.
\end{split}
\end{equation}

\end{document}